\documentstyle[equations,aipconf,times,graphicx]{article}

\advance\topmargin -20mm
\def\br{\penalty50 }

\def\section*#1{\par\vspace{10pt}{\centering\uppercase{#1}\par}\vspace{5pt}}

\def\bounceave#1{\left<#1\right>}
\def\sbounceave#1{\langle#1\rangle}
\def\lave#1{\overline{#1}}
\def\Ptilde{\widetilde P}
\def\Chat{\widehat C}
\def\Ctilde{\widetilde C}
\def\u{{\bf u}}\def\S{{\bf S}}\def\v{{\bf v}}
\let\d=\partial
\def\fm{f_{\rm m}}

\newcounter{axis}
\def\xaxis#1#2#3{\put(0,-.06){\unitlength=1.01\unitlength
\setcounter{axis}{0}%
\begin{picture}(0,0)(0,0)%
\multiput(0,0)(#2,0){#1}{\makebox(0,0)[t]{\arabic{axis}}%
\addtocounter{axis}{#3}}%
\end{picture}}}

\def\yaxis#1#2#3{\put(-.05,0){\unitlength=.98\unitlength
\setcounter{axis}{0}%
\begin{picture}(0,0)(0,0)%
\multiput(0,0)(0,#2){#1}{\makebox(0,0)[r]{\arabic{axis}}%
\addtocounter{axis}{#3}}%
\end{picture}}}
\begin{document}\thispagestyle{empty}
\twocolumn[{\centering
{\baselineskip 10pt{\footnotesize {\it Radio-Frequency Power in Plasmas},
edited by R. McWilliams, AIP Conf.\ Proc.\ vol.\ 190, New York,
(1989), pp. 430--433.  Presented at 8th Topical Conf.\ on
RF Plasma Heating, Irvine, Calif., May 1--3, 1989.}\\[12pt]}
\uppercase{Green's function for rf-driven current in a toroidal plasma}\\[6pt]
Charles F. F. Karney, Nathaniel J. Fisch, and Allan H. Reiman\\
Plasma Physics Laboratory, Princeton University, Princeton, NJ 08543-0451
\\[18pt]}]

\section*{Abstract}
The Green's function for rf-driven currents in  a toroidal plasma is
calculated in the long mean-free-path limit.

\section*{Introduction}

The ``adjoint'' technique provides an elegant method for computing the
current driven in a plasma by externally injected radio-frequency waves.
This technique involves using the self-adjoint property of the linearized
collision operator to express the current in terms of a Green's function,
where the Green's function is proportional to the perturbed distribution in
the presence of an electric field.  This technique was first applied by
Hirshman \cite{Hirshm80} for the case of neutral-beam driven currents in a
homogeneous magnetic field.  The effect of the trapped particles found in
toroidal confinement systems was included by Taguchi \cite{Taguch82}.  The
application of this technique to rf-driven currents was made by Antonsen
and Chu \cite{AntChu82}, Taguchi \cite{Taguch83}, and Antonsen and Hui
\cite{AntHui84}.  Approximations to the Green's function have been given by
a number of authors \cite{AntHui84,YAO86,YosAnt86,Chan87,Giruzz87,Cohen87}.

In this paper, we solve the adjoint problem numerically in toroidal
geometry using the collision operator for a relativistic plasma
\cite{BelBud56,BraKar87,BraKar89}.  The pertinent approximations are: the
device is assumed to be axisymmetric; the mean-free-path is assumed to be
long compared to the device (the ``banana limit''); drifts of the electrons
away from the initial flux surface are neglected; in addition the expansion
of the collision operator in Legendre harmonics is truncated after
the $P_1(\cos\theta)$ term.  By posing the problem in terms of a Green's
function, we are, of course, also assuming that the plasma is close enough
to equilibrium for the collision operator to be linearized, and that the
wave-driven flux is known.

\section*{Basic Equations}

In the long mean-free-path limit, the electron distribution on a particular
flux surface is, to lowest order, a function of the collisionless constants
of motion.  We choose to express the distribution in terms of the
``midplane'' coordinates \cite{KerMcC85} $(u_0, \theta_0)$, the magnitude
and direction with respect to the magnetic field of the momentum per unit
mass (henceforth called just the momentum) at the position where the
magnetic field is minimum.  Measuring position on the flux surface by the
length $l$ along the field line from this point, the momentum $(u,\theta)$
at an arbitrary position is
\begin{displaymath}
  u=u_0, \quad \sin^2\theta= b\sin^2\theta_0,
\end{displaymath}
where $b = B(l)/B(0)$.  Particles with $\sin^2\theta_0>
\sin^2\theta_{\rm tr}=1/b_{\rm max}$ are {\it trapped}; other particles are
{\it passing}.  Assuming that the rf is sufficiently weak, the distribution
satisfies
\begin{equation}\label{fp-direct}
\fm\sbounceave{\Chat(f/\fm)}=
\frac1\lambda\frac{\d}{\d\u_0}\cdot\lambda\S_0,
\end{equation}
where
$\fm\Chat(f/\fm)$ is the collision operator linearized about a
Maxwellian $\fm$,
\begin{displaymath}
   \sbounceave{A}=\frac 1{\tau_b}\int\frac{dl}{v_\parallel}A
\end{displaymath}
is the bounce-averaging operator, 
$\tau_b = \int dl/v_\parallel$
is the bounce time, $\lambda=\tau_b v_0\cos\theta_0/L$, and
$L$ is total length of the field line (from one intersection with the
midplane to the next).  $\S_0$ is the rf-induced flux in momentum space
expressed in midplane coordinates.  This is related to the local rf-induced
flux $\S$ via
\begin{displaymath}
  S_{0u}=\sbounceave{S_u}, \qquad
  S_{0\theta}=\bounceave{S_\theta\frac{\tan\theta_0}{\tan\theta}}.
\end{displaymath}
We should also include a term in eq.~(\ref{fp-direct}) which reflects the
slow heating of the background electrons \cite{AntChu82}.  However, this
term does not contribute to the current carried by $f$.

The power dissipated by the wave between two neighboring flux surfaces is
\begin{equation}\label{W-def}
  W=\frac{L\,dV}{\int dl/b} P_0,
\end{equation}
where $dV$ is the elemental volume between the two surfaces and
\begin{equation}\label{P0-def}
  P_0=m\int d^3\u_0\,\lambda \S_0\cdot\v_0.
\end{equation}
The current density at the midplane is 
\begin{equation}\label{J0-def}
  J_{0\parallel}=q\int d^3\u_0\, v_0\cos\theta_0 f.
\end{equation}
At an arbitrary point the current density is $J_\parallel=b
J_{0\parallel}$.  The total toroidal current flowing between two
neighboring flux surfaces is
\begin{equation}\label{I-def}
  I=\frac{Q\,dV}{ \int dl/b} J_{0\parallel},
\end{equation}
where
\begin{displaymath}
  Q=\int \frac{dl}{2\pi R}\frac{B_\zeta}{B}
\end{displaymath}
is the safety factor, $R$ is the major radius, and $B_\zeta$ is the
toroidal magnetic field.

Rather than determine $J_{0\parallel}$ directly by
solving eq.~(\ref{fp-direct}), we consider the adjoint problem,
\begin{equation}\label{fp-adj}
   \sbounceave{\Chat(\chi)}=
-q \frac{v_0\cos\theta_0}{\lambda}\Theta,
\end{equation}
where $\Theta=1$ for passing particles and $0$ for trapped particles.  This
is the equation for the perturbed electron distribution is the presence of
a toroidal loop voltage $TL/Q$.  The rf-driven current density is then
given by \cite{AntChu82}
\begin{equation}\label{J-adj}
  J_{0\parallel}=\int d^3\u_0\, \lambda \S_0\cdot \frac{\d \chi}{\d\u_0}.
\end{equation}
We will express the current drive efficiency by the ratio $\eta =
J_{0\parallel}/P_0$.  Another useful measure of efficiency is in terms of
the macroscopic variables $I$ and $W$, namely
\begin{displaymath}
  \frac IW = \frac QL \eta.
\end{displaymath}

\section*{Bounce-Averaged Collision Operator}

The linearized collision operator is made up of three terms
\begin{eqalignno}
  \Chat(\chi)&=\bigl(C^{e/e}(\fm\chi,\fm)+C^{e/e}(\fm,\fm\chi)\nonumber\\
&\quad+C^{e/i}(\fm\chi,f_i)\bigr)/\fm,\label{C-hat}
\end{eqalignno}
where $C^{a/b}(f_a,f_b)$ is the collision operator for distribution $f_a$
colliding off distribution $f_b$.  The electron-ion term is computed in the
Lorentz limit (with $m_i\to\infty$).  It can be combined with the first
term and bounce averaged to give
\begin{eqalignno*}
 & \frac{1}{u_0^2}\frac{\d}{\d u_0} u_0^2
     D_{uu}\frac{\d\chi}{\d u_0} + F_u \frac{\d\chi}{\d u_0} \\&+
  \frac{D_{\theta\theta}}{u_0^2}
  \frac{1}{\lambda\sin\theta_0}
     \frac{\d}{\d\theta_0}\sin\theta_0 
   \lambda\bounceave{\frac{\tan^2\theta_0}{\tan^2\theta}}
        \frac{\d\chi}{\d\theta_0},
\end{eqalignno*}
where $D_{uu}$ and $F_u$ are the coefficients of energy diffusion and drag
due to electron-electron collisions, and $D_{\theta\theta}$ is the
pitch-angle scattering coefficient due to collisions with both electrons
and ions.  These are given by one-dimensional integrals over a Maxwellian
distribution \cite{BraKar89}.

Let us now turn to the second term in eq.~(\ref{C-hat}).  Since $\chi$ is odd
in $u_\parallel$, we can expand $\chi(u,\theta,l)$ in
terms of spherical harmonics as follows:
\begin{displaymath}
  \chi(u,\theta,l)=\sum_{k\ \rm odd} \chi_k(u,l) P_k(\cos\theta),
\end{displaymath}
where $P_k$ is the Legendre polynomial of degree $k$ and
$\chi_k(u,l)=(2k+1)\int_0^{\pi/2} \chi(u,\theta,l) \br
P_k(\cos\theta)\*\sin\theta\,\br d\theta$.  The linearized collision
operator is a spherically symmetric, so that its angular eigenfunctions are
spherical harmonics.  This allows us to write the term $C^{e/e}(\fm,\br
\chi_k(u,l)\br P_k(\cos\theta))/\br \fm$ as $\Ctilde_k  \equiv 
P_k(\cos\theta)\br
I_k(\chi_k(u,l))$, where $I_k$ is a linear integral
operator.  Transforming to midplane coordinates, we find
\begin{displaymath}
  \chi_k(u,l)=(2k+1) b \int_{0}^{\pi/2}
  \chi(u_0,\theta_0) \Ptilde_k \sin\theta_0\,d\theta_0,
\end{displaymath}
where 
$\Ptilde_k = P_k(\cos\theta){\cos\theta_0}/{\cos\theta}$ and we have used the
fact that  $\chi$ is zero for trapped particles.
The bounce-averaged collision term becomes
\begin{displaymath}
  \sbounceave{\Ctilde_k}=
  \frac{1}{\lambda}\int\frac{dl}{L} 
  \Ptilde_k
  I_k(\chi_k(u,l)).  
\end{displaymath}
Evaluating these expressions is simplified by decomposing $\Ptilde_k$
into midplane Legendre harmonics:
\begin{eqalignno*}
  \Ptilde_1 &= P_{1,0}, \\
  \Ptilde_3 &= -(b-1)P_{1,0}+b P_{3,0}, \\
  \Ptilde_k &= \sum_{k'=1,3,\ldots}^k G_{k,k'} P_{k',0},
\end{eqalignno*}
where $P_{k,0}=P_{k}(\cos\theta_0)$ and $G_{k,k'}$ is a polynomial in $b$.
The collision term can now be written as
\begin{displaymath}
  \sbounceave{\Ctilde_k} = \frac1{\lambda}
  \sum_{k',k''\ \rm odd}^k
  H_{k,k',k''} P_{k',0} I_k(\chi_{k'',0})\Theta,
\end{displaymath}
where
\begin{displaymath}
  H_{k,k',k''}=\frac{2k+1}{2k''+1}
\lave{ b G_{k,k'} G_{k,k''}},
\end{displaymath}
$\chi_{k,0} = \chi_k(l=0)$, and $\lave A=\int A\,dl/L$.
In particular, we have
\begin{displaymath}
  \sbounceave{\Ctilde_1}
  =\lave{b} \frac{\cos\theta_0}{\lambda} I_1(\chi_{1,0}) \Theta.
\end{displaymath}
At present, we include only the $k=1$ term, ignoring all terms
$\sbounceave{\Ctilde_{k\ge3}}$.   We can estimate the error incurred by
comparing the results we get for the electrical conductivity with those of
Rosenbluth {\it et al} \cite{RHH72}.  This indicates that the relative
error in $\chi$ is on the order of $0.05\sqrt\epsilon$ where $\epsilon$ is
the inverse aspect ratio.

\section*{Numerical Solution}

Putting all the terms in eq.~(\ref{fp-adj}) together, we obtain for the
passing particles
\begin{eqalignno*}
 & \frac{1}{u_0^2}\frac{\d}{\d u_0} u_0^2
     D_{uu}\frac{\d\chi}{\d u_0} + F_u \frac{\d\chi}{\d u_0} \\&+
  \frac{D_{\theta\theta}}{u_0^2}
  \frac{1}{\lambda\sin\theta_0}
     \frac{\d}{\d\theta_0}\sin\theta_0 
   \lambda\bounceave{\frac{\tan^2\theta_0}{\tan^2\theta}}
        \frac{\d\chi}{\d\theta_0}\\&+
\lave{b} \frac{\cos\theta_0}{\lambda} I_1(\chi_{1,0})
+q\frac{v_0\cos\theta_0}{\lambda}=0.\yesnumber\label{numerical}
\end{eqalignno*}
We solve this integro-differential equation numerically in the domain
$0\le\theta_0\le\theta_{\rm tr}$, with boundary condition 
$\chi(\theta_0=\theta_{\rm tr})=0$.

A simple magnetic field configuration 
with circular flux surfaces is chosen.  Designating the poloidal angle by
$\phi$, we choose
\begin{eqalignno*}
R&=R_0(1+\epsilon \cos\phi),\\
B_\zeta&=B_{\zeta0}/(1+\epsilon\cos\phi),\\
B_\phi&=B_{\phi0}/(1+\epsilon\cos\phi),\\
b&=(1+\epsilon)/(1+\epsilon\cos\phi).
\end{eqalignno*}
This gives $l/L=\phi/2\pi$,
$Q=(B_{\zeta0}/B_{\phi0})\*\epsilon/\sqrt{1-\epsilon^2}$,
and $L=2\pi R_0 Q \sqrt{1+B_{\phi0}^2/B_{\zeta0}^2}$.

We normalize velocities and momenta to the thermal speed $u_t=\sqrt{T/m}$,
times to the inverse collision frequency
$\nu_t^{-1}$, where $\nu_t=n q^4\*\log\Lambda/4\pi\epsilon_0^2 m^2 u_t^3$,
$\chi$ to $qu_t/\nu_t$, efficiency $\eta$ to $q/mu_t\nu_t$.  
The plasma is characterized three
dimensionless parameters: $T/mc^2$, $\epsilon$, and the effective ion
charge state $Z$.

Level curves for $\chi(u_0,\theta_0)$ for a typical case are shown in
fig.~\ref{contour}.  In computing the efficiency we specialize to waves
which push the particles parallel to the magnetic field.  First, we
suppose that the wave is absorbed in a single location in momentum space,
i.e., $\S_0\propto\delta(\u_0-\u_{0}')\hat\u_{0\parallel}$.  The efficiency
is given by $\eta = (\d \chi/\d u_{0\parallel})
/v_{0\parallel}$ evaluated at $\u_0'$.  This is shown as a
function of $\u_0'$ in fig.~\ref{effcont}.  This shows where in momentum
space we should try to have waves absorbed in order to maximize the
efficiency.

We consider current drive by waves which are Landau damped.  We assume that
the rays pierce the flux surface at a single poloidal angle $\phi'$ where
$b=b'$, and that the wave does not alter the slope of the electron
distribution appreciably.  In this case, we have $\S\propto \fm
\delta(v_\parallel -v_{\rm ph})\* \delta(\phi-\phi') \hat{\u}_\parallel$
and the bounce-averaged flux is given by $\lambda \S_0\propto \fm
\delta(v_0\cos\theta' -v_{\rm ph}) \hat{\u}_{0\parallel}$, where
$\sin^2\theta' = b'\sin^2\theta_0$.  The current drive efficiency can be
calculated by inserting this form for $\S_0$ into eqs.~(\ref{P0-def}) and
(\ref{J-adj}) and performing the integrals numerically.  The resulting
efficiencies are given in fig.~\ref{landau}.  This confirms 
that toroidal effects
reduce the efficiency of current drive and that this reduction can be minimized
if the waves are absorbed on the high-field side of the torus where there
are fewest trapped particles.

\section*{Acknowledgements}
The authors would like to thank Dave Ehst, Steve Hirshman, and Dieter Sigmar for
enlightening discussions.  This work was supported by the U.S. Department of
Energy under contract DE--AC02--76--CHO--3073.

\vspace{\fill}
\pagebreak

\bibliographystyle{aip}
\bibliography{irvine}

\def\figinsert#1{\put(-0.07,0.98){\includegraphics
[width=1.04\unitlength,height=1.09\unitlength,angle=270]{#1}}}
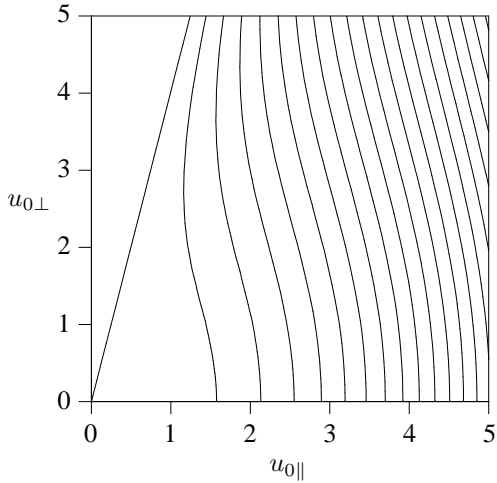
\begin{figure}[ht]
\begin{picture}(1.3,1.3)(-.25,-.2)
\figinsert{adjconhot3}
\put(0.5,-.18){\makebox(0,0)[b]{$u_{0\parallel}$}}
\put(-.1,0.5){\makebox(0,0)[r]{$u_{0\perp}$}}
\xaxis{6}{0.2}{1}\yaxis{6}{0.2}{1}
\end{picture}
\caption[Contour plot]{\label{contour} Contour plot of $\chi(u_0,\theta_0)$
for $Z=1$, $\epsilon=0.03$, and $T/mc^2=0.05$.  The levels of the contours
are given by $\chi=5j$ for integer $j\ge0$ increasing from left to righ.}
\end{figure}

\begin{figure}[ht]
\begin{picture}(1.3,1.3)(-.25,-.2)
\figinsert{effconhot3}
\put(0.5,-.18){\makebox(0,0)[b]{$u_{0\parallel}'$}}
\put(-.1,0.5){\makebox(0,0)[r]{$u_{0\perp}'$}}
\xaxis{6}{0.2}{1}\yaxis{6}{0.2}{1}
\end{picture}
\caption[Contour plot of efficiency]{\label{effcont} Contour plot of
efficiency $\eta$ of current drive with point excitation of
the rf, i.e., $\S_0\propto\delta(\u_0-\u_{0}')\hat\u_{0\parallel}$.  Here,
$Z=1$, $\epsilon=0.03$, and $T/mc^2=0.05$.  The levels of the contours are
given by $\eta=2j$ for integer $j\ge1$ increasing from the origin
outwards.}
\end{figure}
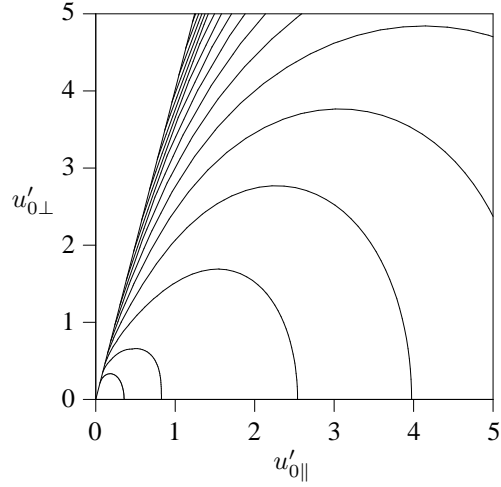

\begin{figure}[ht]
\begin{picture}(1.3,1.3)(-.25,-.2)
\figinsert{adjpltahot2}
\put(0.625,-.15){\makebox(0,0)[b]{$v_{\rm ph}$}}
\put(-.15,0.625){\makebox(0,0)[r]{$\eta$}}
\xaxis{3}{0.5}{1}\yaxis{5}{0.25}{5}
\put(0.55,0.40){\makebox(0,0)[b]{$\epsilon=0$}}
\put(0.75,0.26){\makebox(0,0)[t]{$\epsilon=0.03$}}
\put(0.17,0.15){\makebox(0,0)[l]{$\phi'=0$}}
\put(0.12,0.25){\makebox(0,0)[r]{$\frac12 \pi$}}
\put(0.09,0.45){\makebox(0,0)[tr]{$\pi$}}
\end{picture}
\caption[Landau efficiency]{\label{landau} Efficiencies for current drive
by Lan\-dau-damped waves for $Z=1$, and $T/mc^2=0.05$.  The top curve gives
the efficiency for the case of a uniform magnetic field $\epsilon=0.0$.
The other curves are for $\epsilon=0.03$ and three different poloidal
angles $\phi'$ at which the wave is absorbed.}
\end{figure}
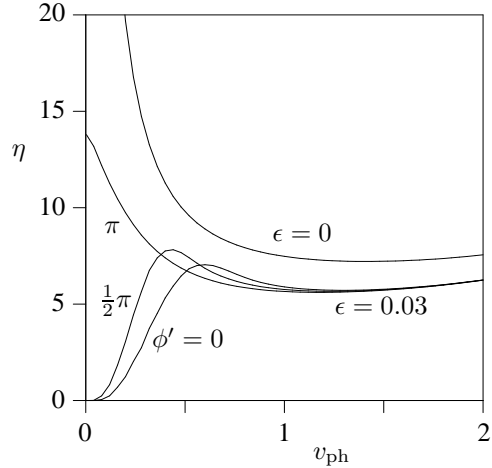

\end{document}